\documentclass[letters,usenatbib]{mnras}
\usepackage{newtxtext,newtxmath}
\usepackage[T1]{fontenc}

\DeclareRobustCommand{\VAN}[3]{#2}
\let\VANthebibliography\thebibliography
\def\thebibliography{\DeclareRobustCommand{\VAN}[3]{##3}\VANthebibliography}

\usepackage{graphicx}	
\usepackage{amsmath}	
\usepackage{xcolor}
\usepackage{multirow}
\usepackage{colortbl}

\newcommand{\Msun}{\rm M_\odot}
\newcommand{\Mvir}{\rm M_{\rm vir}}
\newcommand{\rvir}{\rm r_{\rm vir}}
\newcommand{\rs}{r_S}

\title[Observed vs Simulated Halo c-$\Mvir$ Relations]{Observed versus Simulated Halo 
c-$\Mvir$ Relations}

\author[D. Leier,  et al.]{
Dominik Leier$^{1}$\thanks{E-mail: dominik.leier@gmail.com},
Ignacio Ferreras$^{2,3,4}$, Andrea Negri$^{2,4}$, Prasenjit Saha$^{5}$\\
$^{1}$Dipartimento di Fisica e Astronomia, Alma Mater Studiorum Universit\`{a} di Bologna,
  Viale B. Pichat 6/2, 40127, Bologna, Italy\\
$^{2}$Instituto de Astrof{\'i}sica de Canarias, Calle V{\'i}a L{\'a}ctea s/n,
E38205, La Laguna, Tenerife, Spain\\
$^{3}$Department of Physics and Astronomy, University College London, London WC1E 6BT, UK\\
$^{4}$Departamento de Astrof\'\i sica, Universidad de La Laguna, E38206 La Laguna, Tenerife, Spain\\
$^{5}$Physik-Institut, University of Z\"urich, Winterthurerstrasse 190, CH-8057 Z\"urich, Switzerland\\
}

\date{Accepted 2021 November 09. Received 2021 November 09; in original form 2021 May 05}

\pubyear{2021}

\begin{document}
\label{firstpage}
\pagerange{\pageref{firstpage}--\pageref{lastpage}}

\maketitle

\begin{abstract}
The concentration - virial mass relation is a well-defined trend that
reflects the formation of structure in an expanding
Universe. Numerical simulations reveal a marked correlation that
depends on the collapse time of dark matter halos and their subsequent
assembly history. However, observational constraints are mostly
limited to the massive end via X-ray emission of the hot diffuse gas
in clusters.  An alternative approach, based on gravitational lensing
over galaxy scales, revealed an intriguingly high concentration at
Milky Way-sized halos. This letter focuses on the robustness of these
results by adopting a bootstrapping approach that combines stellar and
lensing mass profiles. We also apply the identical methodology to
simulated halos from EAGLE to assess any systematic. We
bypass several shortcomings of ensemble type lens reconstruction
and conclude that the mismatch between observed and simulated
concentration-to-virial-mass relations are robust, and need to be
explained either invoking a lensing-related sample selection bias, or
a careful investigation of the evolution of concentration
with assembly history. For reference, at a
halo mass of $10^{12}$M$_\odot$, the concentration of observed lenses is
c$_{12}$$\sim$40$\pm$5, whereas simulations give
c$_{12}$$\sim$15$\pm$1.
\end{abstract}




\section{Introduction}

The first stage in the formation of galaxies and clusters involves 
the decoupling of an overdensity from the expanding Universe,
followed by gravitational collapse of that overdensity.  The result is
a halo whose density profile is steep in the outer part but shallow in
the central region. The best-known representative functional form is
the NFW profile \citep{NFW:97} which has logarithmic slope
($\gamma\equiv d\log\rho/d\log r$) of $-1$ in the central region and
$-3$ in the outer region.  The ratio between the radius enclosing the
total, virialised extent of the halo ($\rvir$) and the scale radius
that characterizes the slope transition ($\rs$), taken as the place
where the slope is isothermal, $\gamma(\rs)=-2$, is defined as the
concentration of the halo: $c\equiv \rvir/\rs$. Simple arguments
invoking the gravitational collapse of small structure and their
subsequent virialisation states that the concentration of dark matter
halos scale with the density of the background at the time of
collapse, so that halos that virialise early should have higher
concentration. This theoretical argument is backed by numerical
simulations that reveal a marked correlation between the virial mass
($\Mvir$ defined as the mass contained within $\rvir$) and the
concentration: due to the bottom-up formation paradigm, more massive
halos assemble, on average, at later times, thus featuring lower
concentrations than those corresponding to lower mass halos
\citep[see, e.g.,][]{Maccio_2007}. This effect is expected from
simple principles involving gravity in an expanding Universe,
and theoretical work has explored in detail the
  dependence of concentration on halo mass, collapse redshift, cosmic
  time, morphology, environment, baryon content, etc \citep[see, e.g.][]{Prada:12,DM:14,DK:15,Correa:15,Klypin:16,Shan:17,Chua:19,Wang:20}.
  However,
a detailed, unambiguous confirmation of the c-M relation from
observational evidence remains elusive due to the fact that dark
matter can only be ``observed'' indirectly through dynamical effects
in visible matter, namely the stellar and gaseous components of
galaxies.

Over large scales, involving galaxy clusters (M$>10^{14}\Msun$), it 
is possible to constrain the dark matter halo properties through the X-ray emission of the intracluster medium \citep{Buote_2007,Ettori:10}, via gravitational lensing \citep{Mandelbaum:08,Merten:15}, as well as 
dynamically, with the projected phase space diagram \citep{Biviano:17}. All methods suggest relatively low halo concentrations (c$\sim$5--10), and a scaling relation that can 
be described as a power law of the halo mass, namely:
\begin{equation}
c=\frac{c_{14}}{1+z} \left( \frac{\Mvir}{10^{14} h^{-1} \Msun} \right)^\alpha  
\label{eq:cm}
\end{equation}
At smaller scales, it is still possible to make use of gravitational
lensing -- in the strong regime -- to constrain the underlying halos, although the
constraints become less robust as the lensing data probes comparatively smaller
regions with respect to cluster-scale halos. It is also possible to model
the halo properties by fitting rotation curves of disc galaxies \citep{Martinsson:13}, or via more generalised dynamical modelling \citep[e.g.,][]{WP:11}.
The observational studies agree on a general anti-correlation between concentration
and virial mass \citep[see, e.g.,][]{Buote_2007,Comerford_2007,LFS12}, including results based on non-parametric 
approaches to the density profile \citep{LFSF:11}, however, there are quantitative differences in the scaling law. The origin of these discrepancies may lie in the basic 
methodological requirements and assumptions or in strongly varying sample selections. 
Sample differences originate in substantially different halo environments, feedback 
mechanisms or merger histories to mention a few. It is therefore all the more 
important to apply same methodological approaches to the different halo samples as 
far as possible. Furthermore, over galaxy scales, baryon-related
processes such as adiabatic contraction \citep{Gnedin:04} operate more
efficiently than over larger scales, so a comparison of halo
properties covering a wide range of mass allows us to understand the
baryon physics that regulates galaxy formation.

\begin{figure}
  \includegraphics[width=\columnwidth]{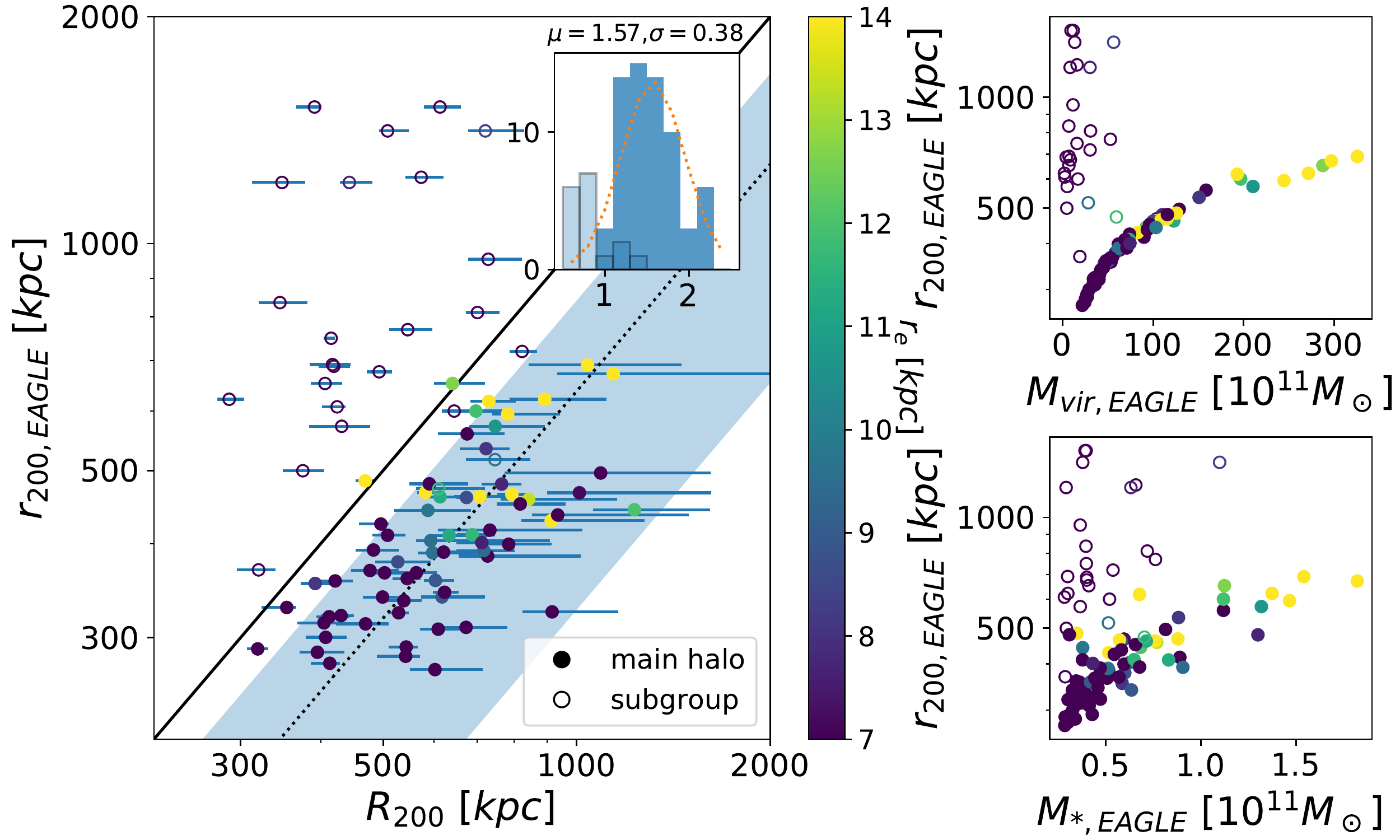}
  \caption{Left panel: Comparison of $R_{200}$ for EAGLE halos calculated using
    the extrapolation method described in Section~\ref{sec:method}
    with $r_{\rm 200,EAGLE}$ inferred by direct extraction from the
    simulation data. The error bars show the $90\%$ CI of the 300
    realizations for each halo taking into account different stellar
    masses and projection angles. The effective radius $r_e$ is colour
    coded. Open circles mark subhalos or groups which cannot be considered isolated,
    whereas filled circles mark main halos. The solid line marks a 1:1 correspondence. The inset plot shows the distribution of the ratio
    $R_{200}/r_{\rm 200,EAGLE}$, including the equivalent Gaussian
    fit. For main halos only we get a mean ratio of $1.57\pm 0.38$, which is hence included in the scatter plot as a dotted line and shaded area. Adding subgroups would give a mean ratio of $1.33 \pm 0.54$.
     The histogram for the subhalos is highlighted
      by contoured semi-transparent bars. Top right panel: Comparison of $r_{\rm 200,EAGLE}$
      with $M_{\rm vir}$ from the EAGLE simulation.
      Bottom right panel: as in the top panel but for
      stellar mass instead.}
    \label{fig:rr}
\end{figure}

This Letter reports, from a study of 18 well-characterised strong
lensing galaxies and 92 comparable simulated dark-matter halos, that
the well-known concentration virial-mass relation for clusters appears
to extend down to galaxy masses. However, the concentrations obtained
are substantially higher than those predicted from numerical
simulations. For instance, at a virial mass of $10^{12}\Msun$, the
derived concentration is $c\sim 50$, whereas the simulations suggest
$c\sim 10$ \citep{Maccio_2007}. The discrepancy is statistically
larger than the expected scatter found in the simulations. It could be
expected from a systematic from the standard methodology applied to
lensing data or from a sample selection bias \citep{Sereno:15}.
However, it could also reflect differences in the assembly history of
the targeted galaxies \citep{Wang:20}, although current simulations
would struggle to produce halos with $c\sim50$. There are indications
of this conundrum already in earlier work \citep[see Figure~4 in][]{LFS12},
but without the control sample of simulated halos,  these results 
could not exclude the possibility that the analysis and fitting method
somehow biased the inferred concentrations. By assessing the methodology
with a control sample, namely numerical simulations, we show in this
letter that this result is robust, and warrants further investigation.
Note that we use henceforth capital $R$ for projected 2D radii and lower
case $r$ for 3D radii. Unless otherwise noted we give quantities based
on the virial radius, which is the radius at which the mean enclosed
density equals a certain multiple $\Delta_c$ (not necessarily 200) of the critical density following the definition in \citet{Bryan:98}.

\section{Methodology}
\label{sec:method}

In this section, we briefly describe the methodology, laid out in more
detail in \citet{LFS12,LFS16}, that combines the lensing and stellar
mass estimates to fit the density profile of the dark matter halo.

We start with the lensing mass maps provided by {\sc PixeLens}
\citep{2004AJ....127.2604S}.  These were obtained by identifying
point-like features in the lensed images, and solving the lensing
equations for each system, producing a pixellated version of the
surface mass density on the lens plane.  The essence of {\sc PixeLens}
is to provide a large number of mass distributions in a non-parametric
way that are compatible with the lens information, under a number of
assumptions, enforced to produce realistic solutions. In order to
assess uncertainties, we take 300 such maps from {\sc PixeLens}.  The
extended lensed images are not fitted, which increases the uncertainty
somewhat, but the difference is expected to be small
\citep[cf.][]{2021MNRAS.501..784D}.  The stellar mass maps are
produced by comparing the observed photometry with population
synthesis models, following the same pixellated grid on the lens
plane.

For each lens, we  generate an ensemble of $\sim$$10^4$ dark matter mass 
maps, by subtracting a random member of the stellar map ensemble ($M_*$) from
a random member of the lensing (i.e. total) mass ensemble ($M_L$). In this Monte Carlo procedure,
we make sure unphysical cases -- namely those that produce negative dark matter mass
-- are rejected. Therefore, a profile with low $M_L$ at
small radius can only be combined with low $M_*$ values. This is an
additional constraint not considered in our previous analysis presented in 
\citet{LFS16}. We then fit each dark matter map with a spherically symmetric
NFW profile, projected on the lens plane, from which we extract the concentration
($c$) and virial mass ($\Mvir$). 

\begin{figure}
  \includegraphics[width=\columnwidth]{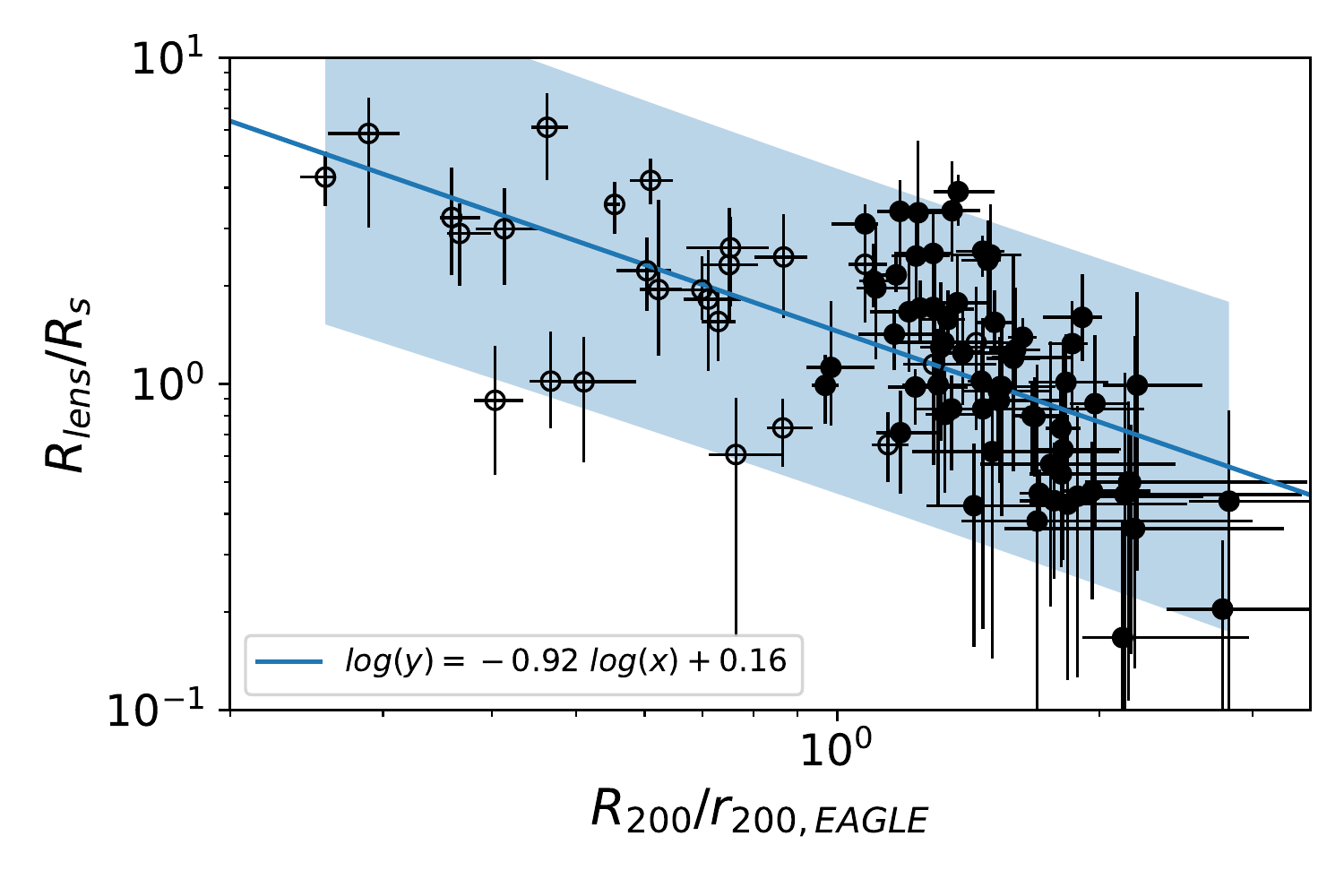}
  \caption{Comparing $R_{lens}/R_s$ with $R_{200}/r_{\rm 200,EAGLE}$ we
    find that the larger the reconstruction radius of the lens with
    respect to the NFW scale radius the lower the virial radius of our
    reconstruction method w.r.t. the $r_{\rm 200,EAGLE}$ taken directly
    from the EAGLE simulation. The blue area shows the $95\%$
    confidence band. As in Fig.~\ref{fig:rr}, we mark main halos (sub halos) with filled (open) circles.}
  \label{fig:rsrr}
\end{figure}

\begin{figure}
\includegraphics[width=\columnwidth]{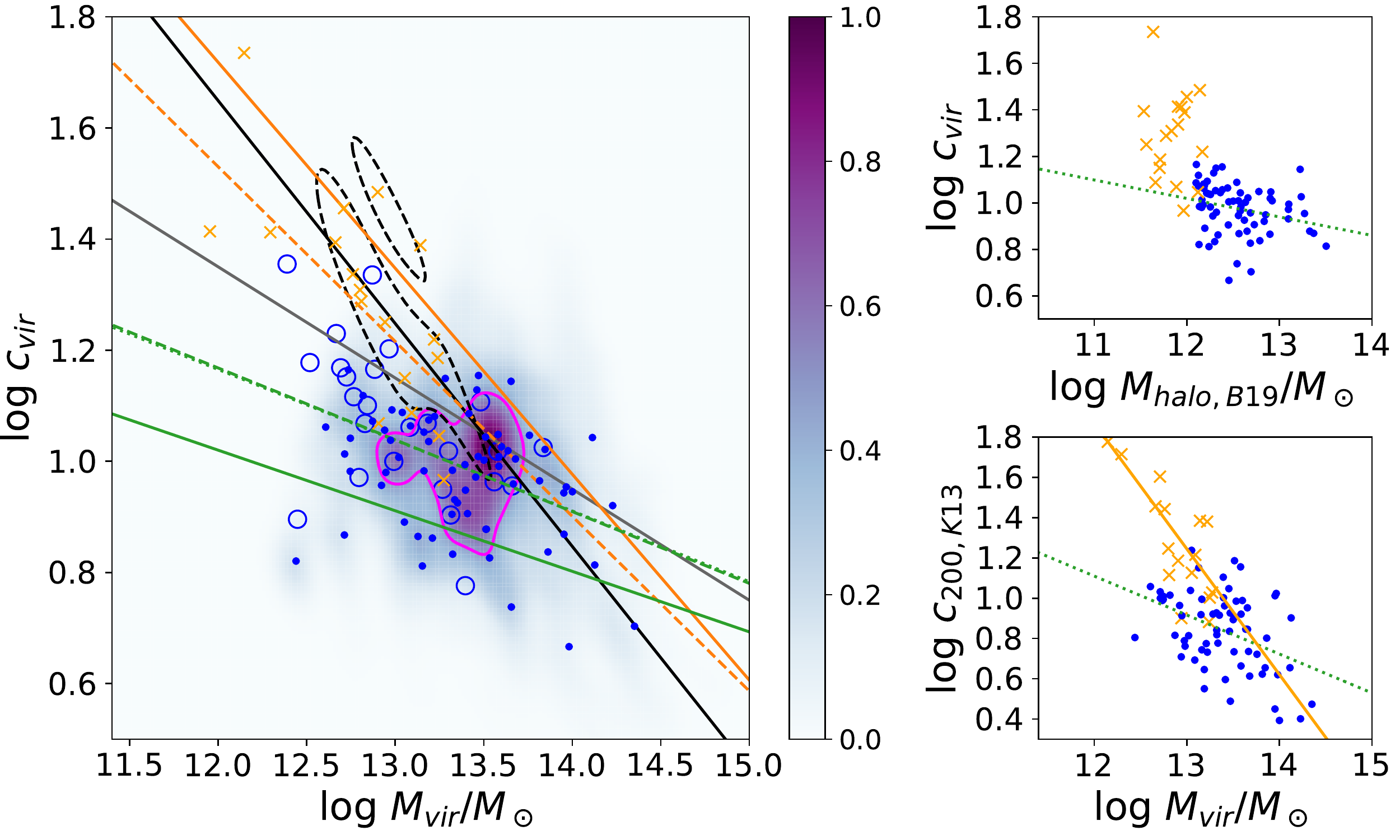}
\caption{Left panel: Concentration vs virial mass plot shown with colour coded
  kernel densities. The solid magenta line shows the $68\%$ density contour in a kernel density estimation of the
  probability distribution for the EAGLE halos with an $\rm RMSD <
  0.06$ corresponding to the median of all RMSD values. The dashed
  black contours show the equivalent for the observed sample of
  lenses. The blue dots (open circles) show the median values of c$_{\rm vir}$ and M$_{\rm vir}$
  of main halos (subhalos) in the EAGLE sample. The orange marks show the sample of observed lenses.
  The solid grey, black and green lines
  are the c-$\Mvir$ relations of the X-ray selected sample of
  \citet{Buote_2007}, our previous best fit \citep{LFS16} and the
  simulation-based relation of \citet{Maccio_2007}, respectively. The
  green dashed line is the best fit to the EAGLE main halo data and the orange solid and dashed
  lines are the best fits to the complete lens sample and the RMSD selected (RMSD < 0.06) sample, i.e. lens$_{<50}$. Top right panel: as in the left panel but with a virial mass inferred by the stellar-to-halo mass relation of \citet{Behroozi:19}. The green dotted line shows the fit to the EAGLE sample. No significant linear regression to the lens sample was found. Bottom right panel: as in the left panel but with a concentration based on the galaxy size-$\rvir$ relation of \citet{Kravtsov:13}.}
    \label{fig:cmfin}
\end{figure}

In addition to the 18 lensing systems presented in \citet{LFS16}, we
explore a set of 92 dark matter halos, selected from the EAGLE
simulation \citep{EAGLE}.  The identical methodology is applied to
these simulated data, to assess potential biases in the derivation of
halo concentration.  The EAGLE data are taken from the $z=0.1$
snapshot of the fiducial simulation RefL0100N1504. We extract the 3D
positional information of the stellar, gas and dark matter particles
considered as part of a given halo. We select all halos with galaxies
more massive than $10^{10.75}\Msun$ in stellar mass.  To account for
halo-to-halo variance, we produce multiple projections of the
simulated halos onto the lens plane by performing 100 random Euler
rotations. The pixel size, Einstein radius and critical mass
corresponding to each set is taken from the observed sample of 18
lenses.  Therefore, the EAGLE sample represents a set with the same
properties as the observed lenses, to remove any sample selection
bias.
Note that, by construction, the simulation data samples the enclosed mass profile at a
radial range comparable with the observed lenses. Combined density contours
weighted by the goodness of fit of each system -- making use of the root mean square deviation,
(hereafter $RMSD$), and the $\chi^2$ statistic -- are then calculated for the
lensing and the simulation data, respectively.

\begin{table}
\begin{center}
\caption{Overview of the $c-\Mvir$ power-law parameters according to
    Eq.~\ref{eq:cm}. In col.~1, B07 refers to the results of \citet{Buote_2007}, M07 to
    \citet{Maccio_2007} and L12 to \citet{LFS12}. The samples ``lens''
    and ``EAGLE'' give the best fits from this paper. If we consider fits
    to our lens sample with a root mean square devation better than the median value
    -- corresponding to the data the densities in Fig.~\ref{fig:cmfin} 
    are based on -- we obtain the results labeled ``lens$_{<50}$''.
    Col.~2 is the slope of the scaling relation, col.~3 is the concentration of the best fit at
    virial mass $10^{14}$M$_\odot$ and col.~4 is the concentration at
    galaxy-scale dark matter halos ($10^{12}$M$_\odot$), the latter shown to
    illustrate the substantially higher concentration of halos in gravitational
    lensing galaxies. We provide errors from bootstrap. See Section~\ref{sec:lensvseagle} for sample details. Note that for lens sample ``$B19$'' no significant fit can be determined. Hence we provide the median concentration.}
\label{tab:cm}
\renewcommand{\arraystretch}{1.1} 
\begin{tabular}{@{}llccc@{}}
\hline
& Sample & $\alpha$ & $c_{14}$ & $c_{12}$\\
& \multicolumn{1}{l}{(1)} & \multicolumn{1}{c}{(2)} & \multicolumn{1}{c}{(3)} & \multicolumn{1}{c}{\hspace{0.2cm}(4)}\\
\hline
\hline
\parbox[t]{2mm}{\multirow{3}{*}{\rotatebox[origin=c]{90}{\colorbox{gray!30}{\hspace{0.5cm}lit.\hspace{0.25cm}}}}} 
& B07   & $-0.199 \pm 0.026$ & $9.12 \pm 0.43$ & $22.80 \pm 1.08$\\ 
& L12   & $-0.401 \pm 0.064$ & $7.03 \pm 1.49$ & $44.56 \pm 9.44$\\
& M07   & $-0.109 \pm 0.005$ & $6.34 \pm 0.26$ & $10.47 \pm 0.43$\\
\hline
\parbox[t]{2mm}{\multirow{4}{*}{\rotatebox[origin=c]{90}{\colorbox{gray!30}{\hspace{0.5cm}lens\hspace{0.5cm}}}}} 
& $all$  & $-0.65 \pm 0.10$ & $3.26^{+1.24}_{-0.90}$ & $65.25^{+7.63}_{-8.64} $\\
& $<50$ & $-0.41 \pm 0.07$ & $6.51^{+1.16}_{-1.41}$ & $43.19^{+5.60}_{-6.43}$\\
& ${B19}$ & $-$ & $-$ & $19.43^{+10.24}_{-6.70}$\\
& ${K13}$ & $-0.62 \pm 0.14$ & $4.18^{+1.29}_{-1.86}$ & $73.91^{+17.50}_{-22.93}$\\
\hline
\parbox[t]{2mm}{\multirow{4}{*}{\rotatebox[origin=c]{90}{\colorbox{gray!30}{\hspace{0.22cm}EAGLE\hspace{0.22cm}}}}} 
& $all$ & $-0.13 \pm 0.03$ & $8.17^{+0.34}_{-0.36}$ & $14.58^{+1.10}_{-1.19}$\\
& ${ISO}$ & $-0.08 \pm 0.03$ & $8.38^{+0.35}_{-0.36}$ & $12.26^{+1.12}_{-1.23}$\\
& ${B19}$ & $-0.08 \pm 0.04$ & $7.23^{+0.81}_{-0.91}$ & $10.43^{+0.31}_{-0.32}$\\
& ${K13}$ & $-0.19 \pm 0.05$ & $5.28^{+0.37}_{-0.39}$ & $12.86^{+0.47}_{-0.49}$\\
\hline
\end{tabular}
\end{center}
\end{table}

Extrapolating the density profile estimate to a certain radius based
on a fit to just a few supporting points at small radii comes along
with uncertainties, which we are able to determine by means of the
EAGLE halos. An underestimate of $R_{200}$, i.e. the radius for which the halo density equals 200 times the critical density, with respect to the actual value, is expected in cases where the fitting points are not
sensitive to the scale radius $R_s$.
To demonstrate the validity of the
extrapolation to the virial radius, $R_{vir}$, and $R_{200}$ respectively, we compare in Fig.~\ref{fig:rr} the EAGLE values of $R_{200}$ with those derived independently, following our methodology. Note that the
uncertainties on the ordinate, i.e. $r_{\rm 200,EAGLE}$, are not shown
to avoid crowding, but are similar to those of our estimates of
R$_{200}$ (abscissa). 
Fig.~\ref{fig:rr} shows that the $R_{200}$ values derived by the above method differ on average by
$57\%$ from the original EAGLE values ($r_{200,EAGLE}$) with the 1:1 correspondence
(solid line) lying $\sim 1.5\sigma$ apart if we consider main halos (filled circles) only. Subhalos (open circles) tend to have systematically larger EAGLE values. We like to point out that if our objects resemble more closely a subgroup environment rather than main halo environments a 1:1 relationship would fall well within $1\sigma$. After all a one-to-one mapping of our lens environments to given definitions of main halos versus subgroups from simulations might be an oversimplification, noting that subhalos themselves may be acting as lenses. However,
for the main halos, a clear correlation between $R_{200}$ and $r_{200,EAGLE}$ is evident. The right top and bottom panels highlight the influence of the environment on $r_{\rm 200,EAGLE}$ and enclosed stellar and virial mass. By analysing the mass enclosed in the range between $r_{\rm 200,EAGLE}$ and the virial radius for the main halos, i.e. $\Mvir-M_{200,EAGLE}$, a quantity that should be sensitive to the mass density in the environment, we find that there is no significant trend with $R_{200}$, $r_{\rm 200,EAGLE}$ or the stellar mass. The small scatter for main halos in the top and bottom panel indicates that EAGLE halos behave largely like the halos in abundance matching relations, stellar-to-halo mass as well as size-$R_{vir}$ relations, which will be further analysed in Section~\ref{sec:lensvseagle}.
More importantly, the uncertainties of $\sim57\%$ (for main halos only) or $\sim33\%$ (for main halos and subgroups) do not explain (a) the large offset
between the concentrations found in the observed lenses and the EAGLE
halos, and (b) the increasing offset towards lower $\Mvir$.  
There is however a significant inverse trend of $R_{200}/r_{200,EAGLE}$ with
$R_{lens}/R_s$ shown in Fig.~\ref{fig:rsrr} where $R_s$ denotes the
NFW scale radius. This implies that our extrapolation will be closer
to the actual values if the reconstructed portion of the lens is large
enough to be sensitive to the radius where the dark matter density
profile turns over to a $r^{-3}$ dependence. However, balancing this
effect by dividing $R_{200}$ by 1.57 for main halos only (1.33 for main halos and subgroups) leads to roughly a $36\%$ ($25\%$) decrease in concentration, which is not enough to reconcile the lensing galaxies with total mass below $10^{13} M_\odot$ with simulations.

There seems to be no obvious correlation with the distribution of
stellar matter as shown by the colour coded effective radii in
Fig.~\ref{fig:rr}. Note that we compute the $90\%$ C.I. uncertainties
of $R_{200}$ by using 100 different projection angles. The
uncertainties of $r_{200,EAGLE}$ (i.e. on the vertical axis) are
roughly of the same extent as they describe halo properties such as
ellipticity and environment. We omit them to avoid crowding in the
plot. In a follow-up study (in preparation) we systematically explore
the dependency of the $c$-$\Mvir$ relation on additional factors such
as the slope of a generalized NFW dark matter profile, the slope of
the IMF, additional matter components as well as variations of the
original lens profiles, to assess various ways of reconciling
observed and simulated data.

\section{Contrasting lens and EAGLE data}
\label{sec:lensvseagle}
A comparison of the sample of 18 observed lenses along with 92
simulated halos from EAGLE is shown in Fig.~\ref{fig:cmfin}. The
lenses cover a range in terms of lens mass, i.e. roughly the total mass
enclosed within the Einstein radius, of $(0.06 - 0.45) \times 10^{12}
\Msun$ and according to our extrapolation up to $R_{vir}$, the sample corresponds to a
virial mass range of $(0.32 - 17.8) \times 10^{12} \Msun$. The orange
crosses represent the observed lenses, whereas the blue marks
correspond to the EAGLE halos. 67 of which are rather isolated main halos (blue dots),
the remaining 25 belong to halo subgroups (open circles). Further key parameters of the lenses
can be found in Table~\ref{tab:ov} below and Table~1 and 3 of \citet{LFS16}. In order to balance the
differing number of valid fits (goodness of fit is evaluated here
using $RMSD$, defined above), and MC realisations, we pick
the same number of $(c,\Mvir)$-points from every lens and halo with a
$RMSD < 0.06$, which corresponds to the median of all $RMSD$ values, to
obtain the orange dashed line in Fig.~\ref{fig:cmfin}. Fitting parameters
to the latter sample, labeled ``${\rm lens}_{<50}$'', are provided in
Table~\ref{tab:cm}. A fit to all $(c,\Mvir)$-points, i.e. unfiltered
with respect to $RMSD$, results in the orange solid line and the parameters
labeled ``lens'' in Table~\ref{tab:cm}. We furthermore fix the IMF to
Salpeter, although we note that the choice of a lighter IMF, such as
Chabrier, does not change the results. 
A comparison of ``lens'' with
``lens$_{<50}$'' shows that even the introduced $RMSD$ selection does not
reconcile the $c-\Mvir$ relation with results from numerical simulations.  We
find that the new analysis of the concentration - virial mass relation
for the observed lens sample is hence in good agreement with the fit
proposed by \citet[][solid black line]{LFS12}.
Moreover, a subsample with the best fit results gives a concentration
at Milky-Way sized halos of c$_{12}\sim40\pm 5$, in stark contrast
with the lower concentrations found in the EAGLE simulations, 
c$_{12}\sim 15\pm 1$. Furthermore, the EAGLE halos -- studied in exactly the same way, 
using the same
pixelisation as the observational sample -- follow a significantly
{\sl shallower} $c-\Mvir$ relation (green dashed line). 
Note that applying the balancing factor of $R_{200}/r_{200,EAGLE}$ ($1.33$ for main and subgroups) from the previous section to both lens and EAGLE sample concentrations, will change c$_{12}$ to $30$ for the lens sample and $\sim 11$ for the EAGLE sample. Also the virial masses will change by a factor of $\sim 2.3$ towards lower values, but still keeping the two c-M relations significantly apart. This is also true if we compare the distributions of concentrations instead of the normalisation c$_{12}$. For the given lens sample we obtain a median concentration of $19.9_{-9.1}^{+14.2}$ versus $9.8_{-3.3}^{+4.0}$ for the EAGLE sample. Applying the above corrections will certainly bring the two populations closer together, but leaves the populations incompatible within their $90\%$ C.I.
In Tab.~\ref{tab:cm} we give the fit parameters for the whole EAGLE sample ('all') and for isolated main halos only ('ISO'). Both fits agree, however, within their uncertainties.
The concentration of the
simulated halos mostly fall below the $c-\Mvir$ relation found by
\cite{Buote_2007} (grey) and the one of \cite{LFS12} (black) -- a finding consistent
with other $c-\Mvir$ relations from simulations. We include the one from 
\citet{Maccio_2007} as a solid green line in Fig.~\ref{fig:cmfin}, and list results of
observational and numerical studies alike in Table~\ref{tab:cm}. Note
that the offset between the EAGLE and the 
\citet{Maccio_2007} scaling relations, both based on numerical simulations, can be
explained by the different redshift of the sample, following Eq.~\ref{eq:cm}.
However, the slopes are in good agreement. The results are put into context of known relations from abundance matching.
We use the stellar-to-halo mass relation by \citet[][B19, top right panel in Fig.~\ref{fig:cmfin}]{Behroozi:19}, and the galaxy size-virial radius relation by \citet[][K13, bottom right panel]{Kravtsov:13}, and find that, despite a certain shift of 0.3-0.5\,dex in halo mass for B19, a rather small average shift of 0.1 going from $\log{c_{vir}}$ to $\log{c_{200}}$ for K13 and a slightly increased scatter, the mismatch between lens and EAGLE c-M fits remains. We like to point out that an offset of $\sim +0.3$ dex is already present when comparing EAGLE halo masses with their respective halo masses according to B19.  Using a sample of EAGLE main halos raises the question whether the lens sample is systematically different with respect to its environment. Based on an extensive discussion in Section~3.1 of \citet{LFS16}, we conclude that half of the lenses reside in cluster or group surroundings which they dominate, and the other half has only isolated companions, if any. Thus, no striking contrast is noticeable when comparing the lens and the EAGLE sample.

\section{Conclusions}
This letter focuses on the concentration vs virial mass relation of dark 
matter halos derived from strong lensing data, extending the results of
\citet{LFS12} by a more robust model fitting methodology, and, most
importantly, contrasting the results from the observed data with 
synthetic systems extracted from the EAGLE cosmological hydrodynamical
simulations, randomised to remove individual variations from halo to halo, 
from which we produce test mass maps in the same format as those used to
explore the observed sample comprising 18 lenses from the CASTLES survey. 
The comparison with simulated data allows us to assess whether the high 
concentrations derived in \citet{LFS12} are real, or rather, produced by
the methodology.

On a concentration-virial mass diagram (Fig.~\ref{fig:cmfin}) we find
the observed halo parameters to be consistent with previous results
from lensing studies and in agreement with X-ray observations within
uncertainties. Furthermore, we show that the difference between the
$c-\Mvir$ relations from numerical simulations and lensing results
cannot be explained by methodological differences as we are able to
recover the shallower slope of the scaling relation known for
simulations while applying the same method to EAGLE halos. Therefore, we
conclude that the high concentrations found in dark matter halos at
virial mass $\sim 10^{12}\Msun$ are not accounted for in
state-of-the-art numerical simulations \citep[e.g.][]{Correa:15,Chua:19}.
Baryonic processes can lead to the contraction of the
  dark matter halo \citep{Blumenthal:86,Gnedin:04}, although the net effect is in
  question, and an
  opposite evolution (i.e. halo expansion) can ensue due to feedback \citep{Dutton:16}. \citet{LFS12} already showed that
  simple adiabatic contraction models cannot account for the mismatch in
  concentration at lower halo mass. The analysis of
  \citet{Chua:19} on Illustris halos shows that the (log) concentration can shift by 
  $\sim+$0.25\,dex, when comparing their full simulations -- including baryon processes --
  with dark-matter only halos.
  Our results suggest that simulated halos {\sl with} baryon physics taken into
  account, are nevertheless underestimating the (log) concentration by as much as
  $\sim+$0.6\,dex in $\sim 10^{12}\Msun$ halos. Even if abundance matching is
  taken at face value, the discrepancy is still $\sim+$0.3\,dex.

The study presented here is part of a series of research efforts that
investigate physical processes and methodological peculiarities such
as sampling biases that may cause the incompatibility of observations
and simulations especially in the low mass range of Early Type
Galaxies. In another paper (Leier et al., in prep.), we will explore in
more detail the bias aspect and peculiarities of the dark matter
profiles of lensing galaxies.

\begin{table}
\begin{center}
\caption{Overview of the lens sample: morphology, scale radius $R_s$, virial radius $R_{vir}$, concentration $c_{vir}$ and mass $M_{vir}$ as shown in Fig.~\ref{fig:cmfin} all with $68\%$ CI. For further lens properties such as stellar mass we refer to Table~3 in \citet{LFS16}. For J1402 and J1719 the probed region of the halos do not constrain the NFW fit by much. Because of this we get many equally good fits. Consequently, introducing a selection based on the median RMSD (see the definition of $lens_{<50}$ in Section~\ref{sec:lensvseagle}) does not help in our effort to narrow down the uncertainties.}
\label{tab:ov}
\begin{tiny}
\renewcommand{\arraystretch}{1.1} 
\begin{tabular}{@{}lllllll@{}}
\hline
Lens ID & Morph.  & $R_{s}$ & $R_{vir}$ & $c_{vir}$ & $M_{vir}$\\
        &         & [kpc]   & [kpc]     &           & [$10^{12}\Msun$] \\ 
\hline
\hline
J0037-0942 & E    & $40.6_{-26.8}^{+43.0}$ & $ 454.57_{-118.7}^{+165.0}$ & $ 11.7_{-4.7}^{+13.0}$ & $ 8.1_{-4.8}^{+12.0}$\\
J0044+0113 & E    & $15.3_{- 3.8}^{+ 6.0}$ & $ 473.1_{- 52.1}^{+ 64.0}$ & $ 30.5_{-5.1}^{+6.0}$ & $ 8.0_{-2.4}^{+4.0}$\\
J0946+1006 & E    & $20.3_{- 5.9}^{+10.0}$ & $ 413.0_{- 34.3}^{+ 52.0}$ & $ 20.4_{-5.1}^{+6.0}$ & $ 6.3_{-1.5}^{+3.0}$\\
J0955+0101 & S    & $37.2_{-22.7}^{+12.0}$ & $ 607.3_{-193.1}^{+ 74.0}$ & $ 16.6_{-2.7}^{+12.0}$ & $ 16.6_{-11.3}^{+7.0}$\\
J0959+0410 & E    & $19.7_{-11.5}^{+20.0}$ & $ 423.2_{-131.5}^{+136.0}$ & $ 21.7_{-7.7}^{+14.0}$ & $ 5.8_{-3.9}^{+8.0}$\\
J1100+5329 & E    & $60.82_{-43.10}^{+30.0}$ & $ 562.7_{-211.7}^{+ 76.0}$ & $ 9.3_{-2.2}^{+11.0}$ & $ 18.8_{-14.2}^{+9.0}$\\
J1143-0144 & E    & $23.8_{-12.6}^{+15.0}$ & $ 573.8_{-167.7}^{+110.0}$ & $ 24.5_{-6.8}^{+12.0}$ & $ 13.9_{-8.9}^{+10.0}$\\
J1204+0358 & E    & $15.9_{- 8.7}^{+19.0}$ & $ 383.8_{-102.1}^{+133.0}$ & $ 24.8_{-9.5}^{+16.0}$ & $ 4.6_{-2.8}^{+7.0}$\\
J1213+6708 & E    & $22.9_{-15.1}^{+28.0}$ & $ 440.0_{-151.1}^{+177.0}$ & $ 19.4_{-7.7}^{+19.0}$ & $ 6.4_{-4.6}^{+11.0}$\\
J1402+6321 & E & $ 4.3_{- 4.1}^{+16.0}$ & $ 252.4_{-66.8}^{+144.0}$ & $ 54.3_{-36.3}^{+72.0}$ & $ 1.4_{-0.8}^{+4.0}$\\
J1525+3327 & E    & $51.6_{-25.0}^{+13.0}$ & $ 539.3_{- 85.3}^{+ 63.0}$ & $ 11.1_{-2.2}^{+6.0}$ & $ 17.7_{-7.1}^{+7.0}$\\
J1531-0105 & E    & $36.0_{-16.4}^{+30.0}$ & $ 519.7_{-104.7}^{+144.0}$ & $ 14.1_{-4.3}^{+6.0}$ & $ 11.3_{-5.6}^{+12.0}$\\
J1538+5817 & E    & $8.70_{-8.3}^{+36.0}$ & $ 225.55_{-93.3}^{+211.0}$ & $ 25.9_{-17.6}^{+272.0}$ & $ 0.9_{-0.7}^{+6.0}$\\
J1630+4520 & E    & $41.8_{-17.9}^{+23.0}$ & $ 509.1_{- 58.1}^{+118.0}$ & $ 12.2_{-3.2}^{+6.0}$ & $ 12.4_{-3.8}^{+11.0}$\\
J1719+2939 & E/S0 & $11.6_{- 9.1}^{+32.0}$ & $ 286.7_{-101.8}^{+185.0}$ & $ 25.8_{-15.5}^{+51.0}$ & $ 2.0_{-1.4}^{+7.0}$\\
J2303+1422 & E    & $27.5_{-13.5}^{+30.0}$ & $ 478.2_{-123.4}^{+145.0}$ & $ 17.81_{-6.6}^{+9.0}$ & $ 8.8_{-5.2}^{+11.0}$ \\
J2343-0030 & E/S0 & $14.2_{- 8.1}^{+23.0}$ & $ 394.5_{-100.2}^{+174.0}$ & $ 28.6_{-13.1}^{+22.0}$ & $ 5.1_{-3.0}^{+10.0}$ \\
J2347-0005 & E    & $33.7_{-12.2}^{+18.0}$ & $ 518.8_{-120.6}^{+70.0}$ & $ 15.4_{-4.5}^{+4.0}$ & $ 17.4_{-9.5}^{+8.0}$ \\
\hline
\end{tabular}
\end{tiny}
\end{center}
\end{table}

\section*{Acknowledgements}

The research of DL is part of the project GLENCO, funded under the
European Seventh Framework Programme, Ideas Grant Agreement
no. 259349. IF is partially supported by grant PID2019-104788GB-I00
from the Spanish Ministry of Science, Innovation and Universities
(MCIU). AN is supported by the Ministerio de Ciencia, Innovaci\'{o}n y Universidades (MICIU/FEDER) under research grant PGC2018-094975-C22. The authors also thank the reviewer for comments which improved the Letter.

\section*{Data Availability}

This sample is extracted from publicly available data from the CASTLES 
sample \citep{CASTLES} and the EAGLE simulations \citep{EAGLE}. The
combined final dataset is available upon reasonable request.



\bibliographystyle{mnras}
\bibliography{references} 


\bsp	
\label{lastpage}
\end{document}